\documentclass[journal=mamobx,manuscript=article,layout=traditional]{achemso}
\SectionsOn
\SectionNumbersOff
\usepackage[version=3]{mhchem} % Formula subscripts using \ce{}
\usepackage{subfigure}

\usepackage{siunitx}
\usepackage{tikz}
\usepackage{nicefrac}
\newcommand{\tikzcircle}[2][black,fill=black]{\tikz\draw[black,fill=black] (0,0) circle (.5ex);}

\usepackage{xcolor}
\usepackage[colorlinks = true,
            linkcolor = blue,
            urlcolor  = blue,
            citecolor = blue,
            anchorcolor = blue]{hyperref}

\author{Shivank S. Shukla}
\author{Christopher Kuenneth}
\author{Rampi Ramprasad}
\email{rampi.ramprasad@mse.gatech.edu}
\affiliation{School of Materials Science and Engineering, Georgia Institute of Technology, Atlanta, Georgia 30332, USA}

\title[Blend Informatics]{Polymer Informatics Beyond Homopolymers}

\usepackage[modulo]{lineno}
\begin{document}

% \makeatletter
% \setlength\acs@tocentry@height{3.5cm}
% \setlength\acs@tocentry@width{8.3cm}
% \makeatother
% \renewcommand\tocentryname{for Table of Contents use only}

% \clearpage
% \section{for Table of Contents use only}
% Title: Polymer Blend Informatics with Multi-Task Deep Neural Networks \\
% Authors: Shivank S. Shukla, Christopher Kuenneth, Rampi Ramprasad \\

% % \begin{figure}[hbt]
% \begin{center}
% \includegraphics{GA_copolymer_informatics.pdf}    
% \end{center}
% \clearpage
% %   \caption{TOC graphic}
% % \end{figure}

% \end{tocentry}

\begin{abstract}
Polymers are diverse and versatile materials that have met a wide range of material application demands. They come in several flavors and architectures, e.g., homopolymers, copolymers, polymer blends, and polymers with additives. Searching this enormous space for suitable materials with a specific set of property/performance targets is thus non-trivial, painstaking, and expensive. Such a search process can be made effective by the creation of rapid and accurate property predictors. In this work, we present a machine-learning framework to predict the thermal properties of homopolymers, copolymers, and polymer blends. A universal fingerprinting scheme capable of handling this entire polymer chemical class has been developed and a multi-task deep learning algorithm is trained simultaneously on a large dataset of glass transition, melting, and degradation temperatures. The developed models are accurate, fast, flexible, and scalable to other properties when suitable data become available.
\end{abstract}

\section{Introduction}

Polymeric materials come in a variety of flavors and architectures, such as homopolymers, copolymers, polymer blends, and polymers with additives such as dopants, plasticizers, and organic/inorganic fillers.\cite{hagiopol1999copolymerization, paul2012polymer, Ambrogi2017, doi:10.1080/00222358008080917} The extraordinary chemical and structural diversity offered by such materials lead to wide-ranging and attractive combinations of physical properties impacting several application spaces, ranging from structural, electrical, packaging, chemical separation, healthcare, energy, and sustainable technologies.\cite{Singh2015a, Bartlett1993a,Nunes2016,Sun2013, ekebafe2011polymer,puoci2008polymer, rasek2009polymers,Robeson1984,Okay2000, Elhefian2014, Yu2006, Potschke2003, Hamad2014}

In an effort to simultaneously optimize multiple (correlated or uncorrelated) properties, the community has explored and developed polymer varieties beyond neat homopolymers, namely, copolymers, polymer blends, and polymers with additives. Finding optimal candidates possessing a predefined set of property attributes has largely been guided by experience, intuition, and trial-and-error approaches. An exhaustive search of the relevant chemical spaces is non-trivial given the vast expanse of the spaces. Over the last decade or so, polymer informatics approaches have attempted to aid this search process by offering data-driven machine learning models to rapidly predict the properties of new polymer formulations and to recommend candidate materials that may meet multi-property target requirements.\cite{Kim2018, Chandrasekaran2020, doi:10.1063/5.0023759, KUENNETH2021100238, Pilania2013, Ramprasad2017, Wu2019a, YAN2021123351, doi:10.1126/sciadv.aaz4301,ZHU2020120381}. Nevertheless, these polymer informatics efforts have thus far largely focused on neat homopolymers,\cite{Chen2021} although notable exceptions exist within recent attempts to address copolymer chemistries,\cite{Kuenneth2021, Zhang2021, Pilania2019, ma13245701, Goswami2021, Zhang2021a} polymers with dopants, and polymer composites. \cite{Zhu2021}

In this contribution, we lay the groundwork to handle neat homopolymers, copolymers, and polymer blends within one unified multi-task neural network polymer informatics framework. For definiteness, we focus on thermal properties, namely, the glass transition temperature ($T_\text{g}$), the melting temperature ($T_\text{m}$), and the degradation temperature ($T_\text{d}$). Once trained (simultaneously) on $T_\text{g}$, $T_\text{m}$, and $T_\text{d}$ data for homopolymers, copolymers, and polymer blends, this new polymer informatics framework can predict these properties for any new homopolymer, copolymer, or polymer blend.

\begin{figure*}[hbt!]
\begin{center}
 \includegraphics[scale=0.95]{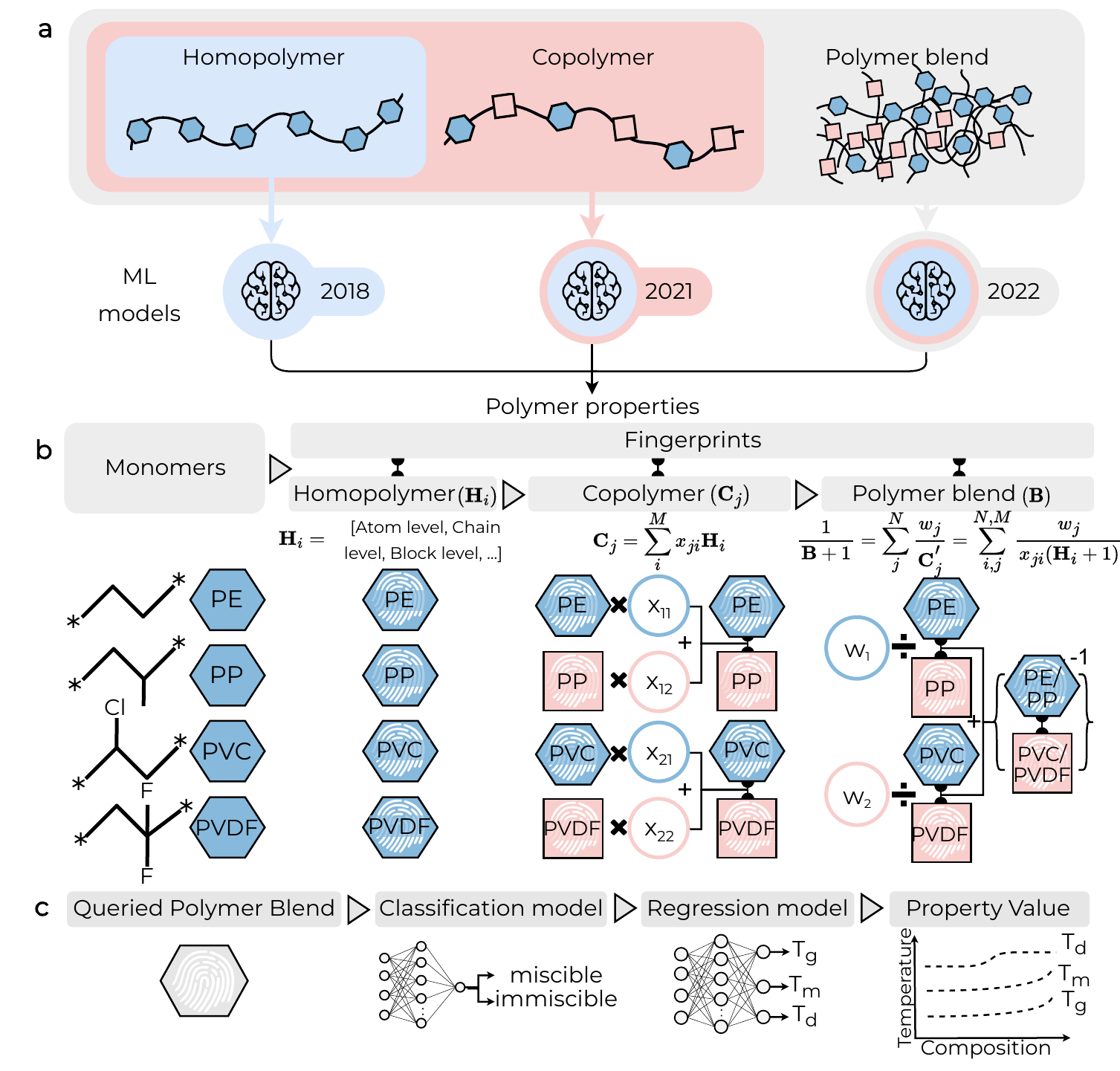}
  \caption{\textbf{a} Evolution of our polymer informatics framework that incorporates predictors for homopolymer \cite{doi:10.1063/5.0023759}, homopolymer and copolymer \cite{Kuenneth2021}, and the current work for homopolymer, copolymer, and polymer blend. \textbf{b} Fingerprint computation pipeline for the example of a poly(ethylene)-co-poly(propylene) and poly(vinyl chloride)-co-poly(vinylidene fluoride) blend. The monomers of poly(ethylene), poly(propylene), poly(vinyl chloride), and poly(vinylidene fluoride) are shown as PE, PP, PVC, and PVDF, respectively. Homopolymer fingerprints ($\mathbf{H}_{i}$) for the homopolymers ($i$) are computed using the polymer genome fingerprinting framework\cite{Kim2018}. Copolymer fingerprints ($\mathbf{C}_{j}$) for the copolymers ($j$) are the composition-weighted sum of the homopolymer fingerprints ($\mathbf{H}_{i}$) with $x_{ji}$ being the compositions of monomer $i$ in copolymer $j$. Polymer blend fingerprints ($\mathbf{B}$) are the composition-weighted harmonic sum of the copolymer fingerprints with $M$, $N$ and $w_{\text{j}}$ representing the number of monomers in copolymer $j$, the number of polymers and their weight fraction in the polymer blend, respectively. \textbf{c} The inference pipeline to predict thermal properties for a new polymer blend.}

  \label{fig:flow_evolution}
  \end{center}
\end{figure*}

As portrayed in Figure \ref{fig:flow_evolution}a, homopolymers are a subset of copolymers, and copolymers are a subset of polymer blends. Homopolymers are defined by one monomer repeat unit and copolymers by multiple monomer units. Polymer blends are a physical mixture of two or more homopolymer(s) and/or copolymer(s). In this work, we assume that our copolymers are random, i.e., the multiple repeat units are distributed randomly along the polymer backbone; we make this assumption because we do not have data that specify the particular archetype of the copolymer. Polymer blends may be miscible or immiscible; our framework first predicts which category the polymer blend belongs to, and then, subsequently, predicts the appropriate number of critical temperatures (miscible polymer blends are characterized by one $T_\text{g}$, one $T_\text{m}$ and one $T_\text{d}$, while immiscible 2-phase polymer blends may display two $T_\text{g}$, two $T_\text{m}$ and one $T_\text{d}$).

Our machine learning procedure starts by first converting the chemical structure of homopolymers, copolymers, and polymer blends into numerical vectors called fingerprints; the chemical structure itself are specified using SMILES strings \cite{Weininger1988} of the repeat units, the composition of the copolymers (if relevant) and the weight fraction of the polymer blend components (if relevant). These aspects are captured in Figure \ref{fig:flow_evolution}b. The fingerprints of the chemical structures, along with $T_\text{g}$, $T_\text{m}$, and $T_\text{d}$ data are fed into our multi-task neural network architecture \cite{Kuenneth2021}, with the ultimate output being a trained model that can predict whether a polymer blend (if the queried case is a polymer blend) is miscible or not, followed by the thermal properties of the queried case. This new polymer informatics capability is able to predict $T_\text{g}$, $T_\text{m}$, and $T_\text{d}$ with an RMSE of \SI{15}{K}, \SI{17}{K}, and \SI{23}{K}, respectively. Needless to say, this framework can be extended to handle any other property class, as long as the requisite data for these properties are available.

\begin{table}[hbtp]
\centering
  \caption{The number of homopolymers, copolymers, and polymer blends data points for the glass transition ($T_\text{g}$), melting ($T_\text{m}$), and degradation ($T_\text{d}$) temperatures. The 7,774 copolymer data points encompass 1,569 distinct copolymer chemistries, ignoring composition information. The 4,573 polymer blend data points consist of 626 distinct combinations of two constituents (two homopolymers, two copolymers, or a homopolymer and a copolymer). Polymer blends can be miscible (M) or immiscible (IM). The reported measurement method, differential scanning calorimetry (DSC), or thermogravimetric analysis (TGA) is also indicated.}
  \label{tbl:data_set}
 
\begin{tabular}{llrrrcc|r}
\hline
% Property & Homopolymer & Copolymer & \multicolumn{2}{c}{Polymer Blend} & Range & Method & Total \\ \hline
Property & Homopolymer & Copolymer & \multicolumn{2}{c}{Polymer Blend} & Range & Method & Total \\ \hline
 % & polymer & polymer & \multicolumn{2}{c}{Blend} &  &  &  \\ \hline
 &  &  & M & IM & & & \\ \hline
$T_\text{g}$ & 5,072 & 4,426 & 2,541 & 440 & {[}131, 587{]} & DSC & 12,039 \\
$T_\text{m}$ & 2,079 & 1,988 & 1,001 & 330 & {[}253, 639{]} & DSC & 5,068 \\
$T_\text{d}$ & 3,520 & 1,360 & 261 & - & $[383, 977]$ & TGA & 5,141 \\ \hline
Total & 10,671 & 7,774 & 3,803 & 770 &  &  & 23,018 \\ \hline
\end{tabular}
\end{table}

\section{Results and Discussion}

\paragraph{Data}

\begin{figure*}[hbt!]
\begin{center}
 \includegraphics{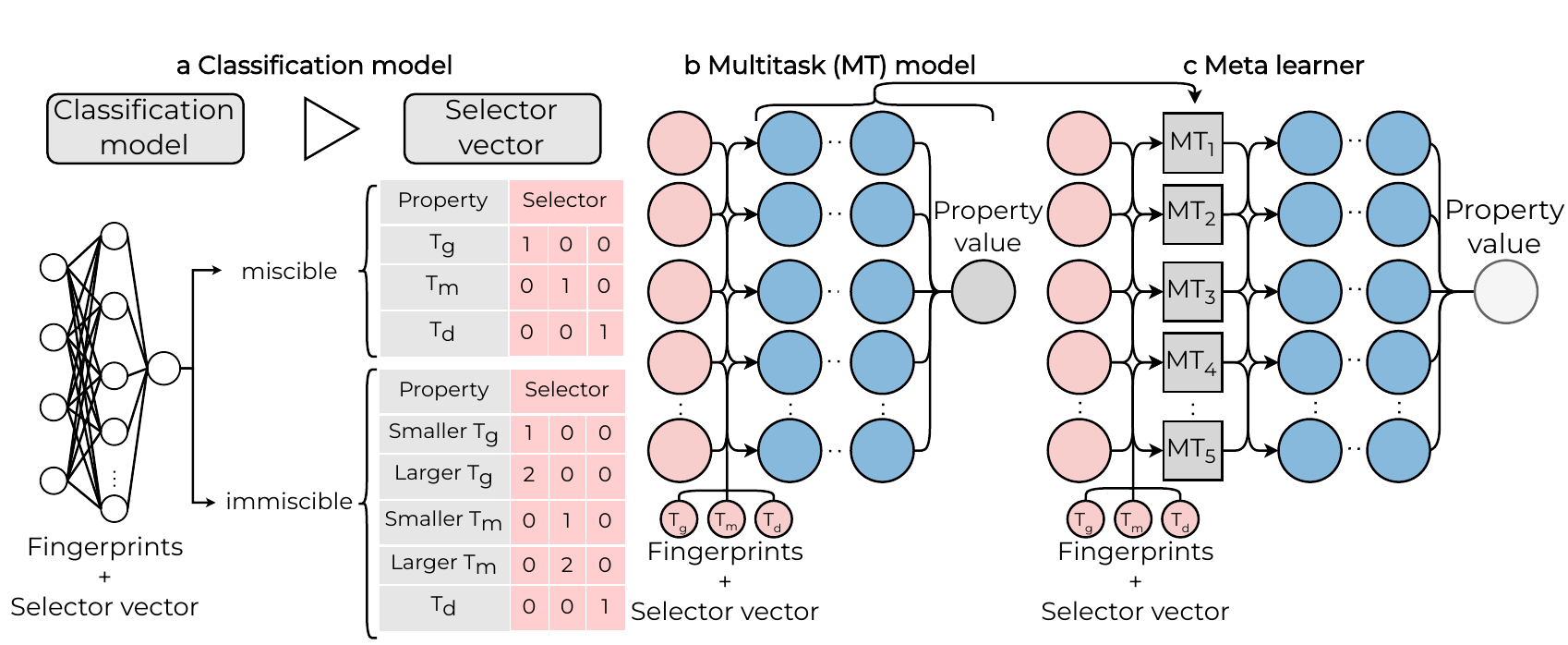}

     \caption{Machine learning workflow. \textbf{a} The multi-task deep neural network based classification model that predicts if a polymer blend is miscible or immiscible and to determine the miscibility-dependent number and components of the selector vectors needed for each thermal property. \textbf{b} The multi-task deep neural network based regression models to predict $T_\text{g}$, $T_\text{m}$, and $T_\text{d}$ with five-fold cross-validation. The inputs to this model are the polymer fingerprints and the selector vector. \textbf{c} The meta learner model. The inputs to this meta learner are the five property values from the five-fold cross-validation models.}

  \label{fig:ml_arch}
  \end{center}
\end{figure*}
 
The dataset used in this study for homopolymer, copolymer, and polymer blends $T_\text{g}$, $T_\text{m}$, or $T_\text{d}$ were collected from multiple sources cited elsewhere \cite{Kim2018,Kim2019,Jha2019a,Kuenneth2021, polyinfo}, including from the polyInfo database \cite{polyinfo} (The copyrights of this database are owned by National Institute for Materials Science (NIMS)). For uniformity and consistency, we only use $T_\text{g}$ and $T_\text{m}$ data points measured using differential scanning calorimetry (DSC), and $T_\text{d}$ values measured via thermogravimetric analysis (TGA). Each copolymer data point has two comonomers, and each polymer blend data point consists of two constituents (two homopolymers, a homopolymer-copolymer mixture, or two copolymers). We infer the polymer blend miscibility for each data point from the presence of one or two $T_\text{g}$ values in the dataset. This is also known as technological miscibility \cite{Manias2014}. If the polymer blend data point has only one $T_\text{g}$ value, we consider it miscible, otherwise, it is classified as immiscible. A three-component selector vector is used to indicate the property ($T_\text{g}$, $T_\text{m}$, or $T_\text{d}$) and miscibility (miscible or immiscible) of the data point. The property (for homopolymers, copolymers, and polymer blends) is encoded in the selector vector at the position of the non-zero component as shown in Figure \ref{fig:ml_arch}a. For polymer blends, the miscibility information is encoded at the appropriate component taking on values of 1 or 2 as shown in Figure \ref{fig:ml_arch}a. 
\begin{figure}[hbt]
\begin{center}
 \includegraphics{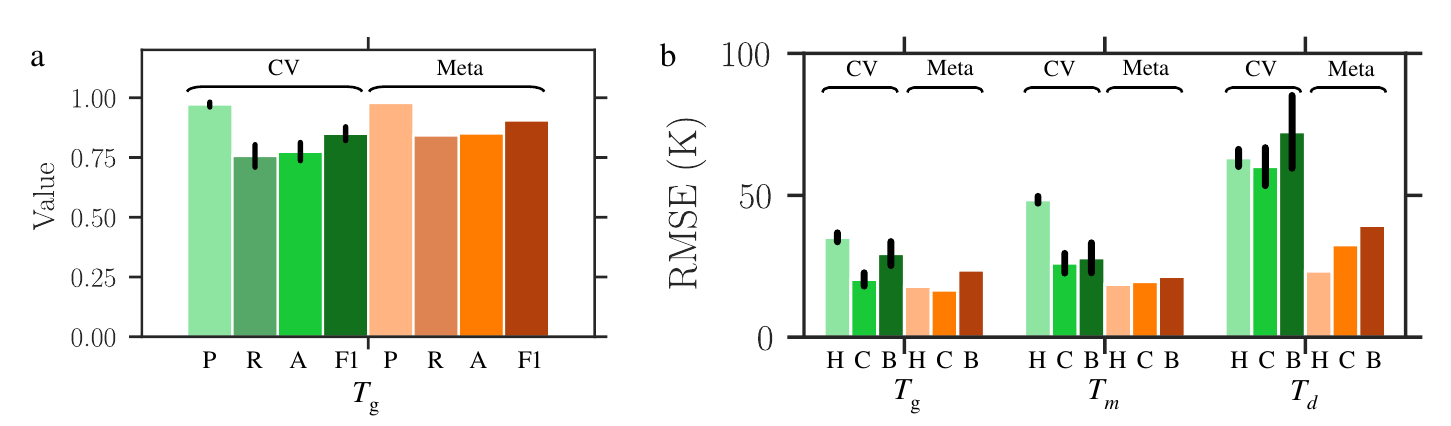}

    % \centering
    % \subfigure[]{\includegraphics[width=0.40\textwidth]{Paper_figure_class.pdf}} 
    % \subfigure[]{\includegraphics[width=0.48\textwidth]{Paper_figure_4.pdf}} 

\caption{$\mathbf{a}$ Validation set performance metrics of the five-fold cross-validation (CV) classification models and meta learner (Meta) model for glass transition ($T_\text{g}$) of polymer blends. The performance metrics are P: precision, R: recall, A: accuracy, and F1: F1 score. The reported values for all these metrics are averaged over the five CV models and the black error bars represent the standard deviation observed in the performance metrics across five CV models. $\mathbf{b}$ Test set root-mean-square error (RMSE) values of five-fold cross-validation (CV) regression models and meta learner (Meta) model for glass transition ($T_\text{g}$), melting ($T_\text{m}$) and thermal degradation ($T_\text{d}$) temperature of homopolymers (H), copolymers (C) and polymer blends (B). For the five-fold CV RMSE, the errors are averaged across the five CV models and the black error bars represent the standard deviation observed in the RMSE values of the five CV models.
}
  \label{fig:performance}
\end{center}
\end{figure}

\paragraph{Fingerprinting}
 
 %\cor{A drawback of fingerprinting in such a manner is the loss of information about the individual constituents in the fingerprinting step. Therefore, we need a scheme that retains this information so that we can compare their predictive performances to draw conclusions. Concatenating the fingerprint vectors of the constituents in the polymer blends with their relative compositions solves the problem of information loss but these fingerprints are permutation dependent with inconsistent length depending on the number of components in the polymer blends. However, for two-component polymer blends, we can exploit the two different arrangements of the stacked fingerprint vector by associating each of them with one of the thermal property values.}

The repeat units of the monomers of homopolymers, copolymers, and polymer blends in the dataset are represented using Simplified Molecular Input Line System (SMILES) strings\cite{Weininger1988}. We use stars \texttt{[*]} to denote the endpoints of the repeat unit. SMILES strings cannot directly be ingested by conventional machine learning models and require conversion to numerical vectors. This conversion is performed using a previously-pioneered handcrafted fingerprinting scheme \cite{doi:10.1063/5.0023759} (see Method section) that has shown great performance for predicting properties of polymers in many previous works \cite{Ramprasad2017, doi:10.1063/5.0023759, Kim2018, Chandrasekaran2020, Chen2021}. For copolymers\cite{Kuenneth2021}, we compute fingerprints as the composition-weighted sum of the homopolymer fingerprint vectors (${\mathbf{C}_{j}} = \sum_{i}^{N} {x_{ji} \mathbf{H}_{i}} $), as shown in Figure \ref{fig:flow_evolution}b. $\mathbf{H}_{i}$, $x_{ji}$, and $N$ denote the fingerprint vector of a homopolymer ($i$), relative compositions of homopolymers in a copolymer ($j$), and the total number of comonomer components (in this work, $N = 1,2$), respectively. For polymer blends, we use the composition-weighted harmonic mean of fingerprint vectors of the constituents in the polymer blend. To compute the polymer blend fingerprints, we use $\nicefrac{1}{\mathbf{B}+1} = \sum_j^M \nicefrac{w_j} {\mathbf{C}^{'}_{j}} = \sum_{i,j}^{N,M}  \nicefrac{w_j} {x_{ji} \mathbf{H}_{i}+1} $. Here, $\mathbf{B}$, $w_{\text{j}}$, $\mathbf{C^{'}_{j}}$ and $M$  represent the polymer blend fingerprint, the relative composition of the $j^{\text{th}}$ constituent, fingerprint vector of the $j^{\text{th}}$ polymer blend constituent and the number of constituents in the polymer blend, respectively. This equation resembles the mathematical form of the Fox equation. \cite{BROSTOW20083152} A scalar factor of one was added to the fingerprint vector components (and later subtracted) to avoid singularities caused by fingerprint components with the value of zero. The complete fingerprinting pipeline for homopolymers, copolymers, and polymer blends is shown in Figure \ref{fig:flow_evolution}b.  

 %\cor{A drawback of fingerprinting in such a manner is the loss of information about the individual constituents in the fingerprinting step. Therefore, we need a scheme that retains this information so that we can compare their predictive performances to draw conclusions. Concatenating the fingerprint vectors of the constituents in the polymer blends with their relative compositions solves the problem of information loss but these fingerprints are permutation dependent with inconsistent length depending on the number of components in the polymer blends. However, for two-component polymer blends, we can exploit the two different arrangements of the stacked fingerprint vector by associating each of them with one of the thermal property values.}
\paragraph{Performance}

ML model development involved five-fold cross-validation (CV) and a meta learner, as described in Methods. The averaged performance scores of the five CV and the meta learner classification models are illustrated in Figure \ref{fig:performance}a. By comparing the precision (P), recall (R), accuracy (A), and F1 score, we find that P is generally higher than R. The reason for this is the imbalance of miscible and immiscible polymer blends (ratio $\approx \nicefrac{5}{1}$) in our dataset (see Table \ref{tbl:data_set}). The performance metrics of the classification meta learner improve from the five-fold cross-validation classification model for $T_\text{g}$ as illustrated in Figure \ref{fig:performance}a. Figure \ref{fig:performance}b and Table S1 (in Supplementary Information) show the root-mean-square error (RMSE) values of the five-fold CV models of the property-predictive regression models. The low RMSE values for $T_\text{g}$, $T_\text{m}$, and $T_\text{d}$ of homopolymers, copolymers, and polymer blends provide confidence in the novel fingerprinting scheme for polymer blends and the usage of multi-task models for this problem. All RMSE values for thermal properties are also of the same order of magnitude as experimental measurement errors. These RMSE values are also slightly better than the performance of homopolymers and copolymers as reported in our past publications. \cite{Chen2021,KUENNETH2021100238,Kuenneth2021} For all types of polymers, RMSEs of five-fold CV are the lowest for $T_\text{g}$, followed by $T_\text{m}$, and then $T_\text{d}$.

%\cline{2-8}
\begin{figure*}[hbt!]
\begin{center}

 \includegraphics{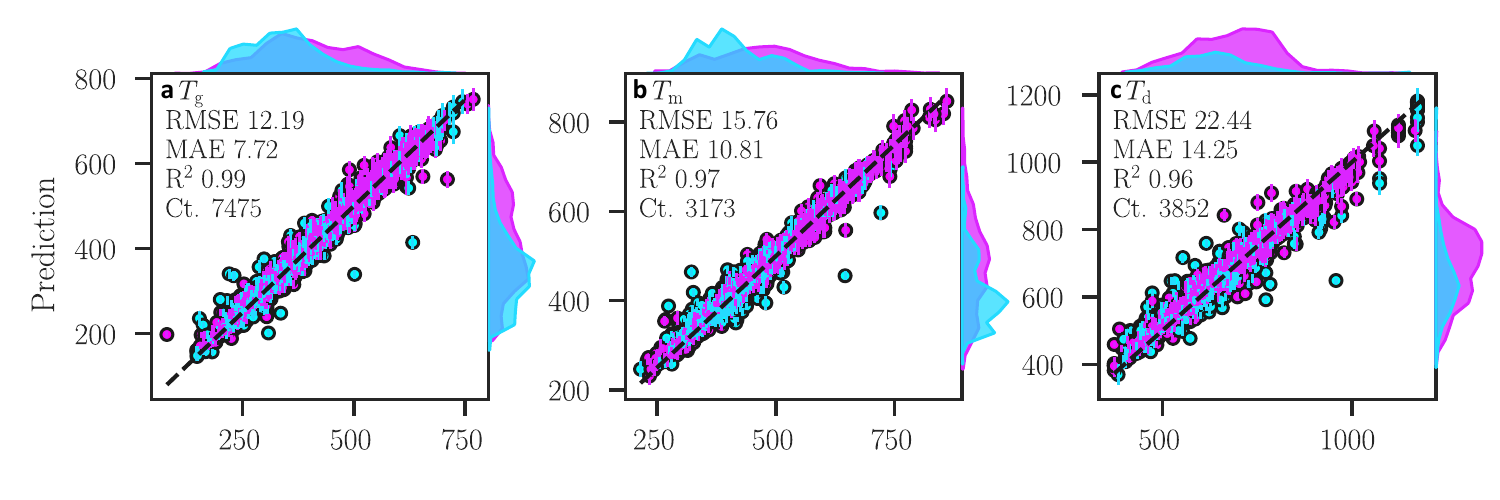}
 \includegraphics{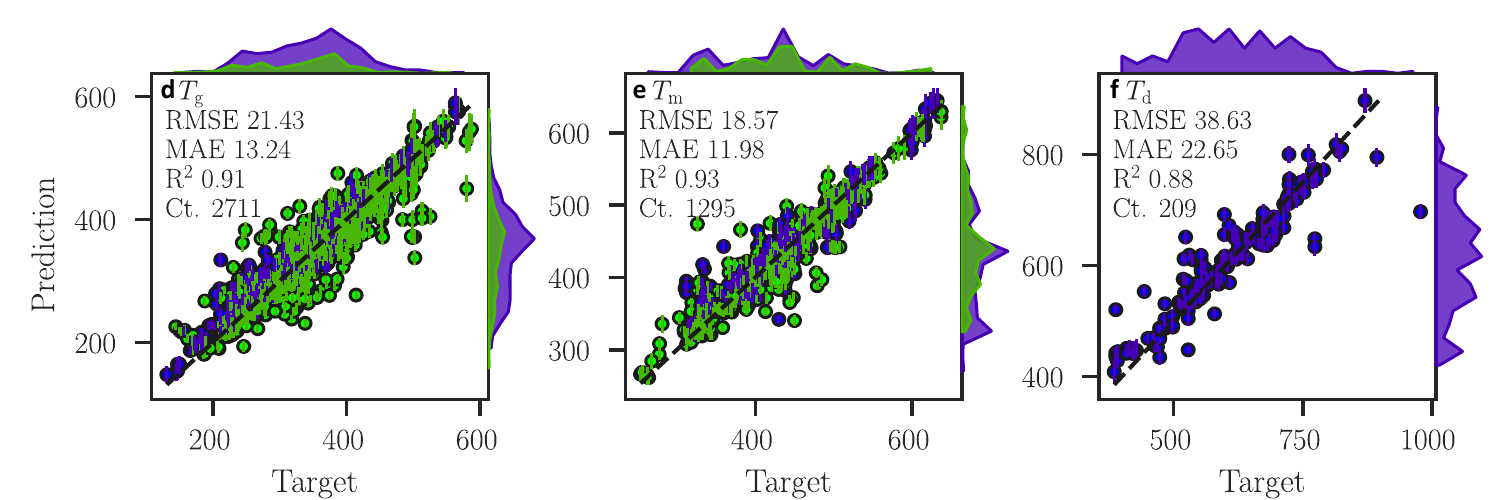}
 \caption{Meta learner parity plots for the test dataset. The first three plots (\textbf{a}-\textbf{c}) are for homopolymers and copolymers represented as pink and cyan data points respectively. The last three plots (\textbf{d}-\textbf{f}) are for polymer blends with one thermal property value and more than one thermal property value (immiscible polymer blends may show two $T_\text{g}$ and two $T_\text{m}$ values) represented as blue and lime data points respectively. The distribution of data points for all properties is shown in the margins of each plot.}
  \label{fig:parity}
  \end{center}
\end{figure*}

The parity plots of meta learner predictions for all the thermal properties on different polymer datasets are shown in Figure \ref{fig:parity}. These meta learner predictions are based on the \SI{80}{\%} dataset used to train the cross-validation models. The low overall RMSE values (including all the types of polymers) of \SI{15}{K}, \SI{17}{K}, and \SI{23}{K}, and high R$^2$ of values 0.98, 0.97, and 0.96 for $T_\text{g}$, $T_\text{m}$, and $T_\text{d}$, respectively, signify high performance across the three distinct classes of polymers. Incorporating a hyperparameter-tuned meta learner on top of the cross-validation model further improves the performance of the multitask model as shown in Table S1 (in Supplementary Information). Monte Carlo dropout is used to estimate the predictions of thermal properties from the meta learner within \SI{95}{\%} confidence interval \cite{Gal2016}. 

\begin{figure}[H]
\begin{center}
 \includegraphics[scale=0.95]{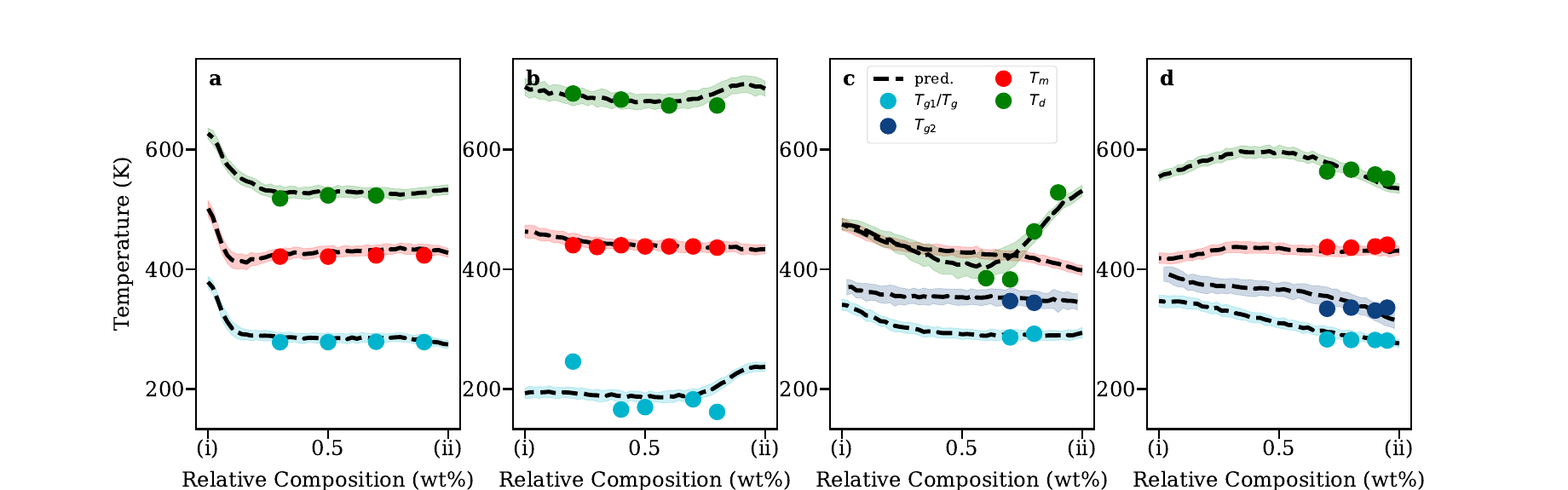}

 \caption{The predicted thermal property values and the experimentally measured thermal property values (solid circles) for $\mathbf{a}$ (i) poly(acrylonitrile-co-styrene) (ii) poly[(3-hydroxybutyric acid)-co-(3-ethyl-3-hydroxypropionic acid)] blend $\mathbf{b}$ (i) poly(vinylidene fluoride) (ii) poly-tetrafluoroethylene-alt-ethylene blend $\mathbf{c}$ (i) poly(propylene carbonate) (ii) poly[(vinyl alcohol)-co-(vinyl acetate)] blend $\mathbf{d}$ (i) poly(oxiranylmethyl methacrylate) (ii) poly(3-hydroxybutyric acid) blend. 
 }
 
%  Plots e), f), g) are ternary prediction plots of poly(vinylidene flouride) and poly[(vinyl acetate)-co-(trifluoroethene)] blend for $T_\text{g}$, $T_\text{m}$ and $T_\text{d}$ respectively. The axes $z_{1}$, $z_{2}$ and $z_{3}$ represent the monomer composition of vinylidene fluoride, trifluoroethene and vinyl acetate within the polymer blend respectively. Plots h), i), j) are the surface prediction plots with the experimental values (solid circles) of poly(vinylacetate)-co-poly(vinylidene fluoride) and poly(vinylacetate)-co-poly(trifluoroethene) blend for $T_\text{g}$, $T_\text{m}$ and $T_\text{d}$ respectively.}
%  \includegraphics[scale=0.75]{Paper_figure_1b.drawio.pdf, }
%  \includegraphics[scale=0.75]{Paper_figure_1c.drawio.pdf}
  
  \label{fig:predictions}
  \end{center}
  
\end{figure}

\paragraph{Sample Prediction}
Model predictions and experimentally measured values across the whole composition range for two selected miscible (Figures \ref{fig:predictions}a and \ref{fig:predictions}b) and two immiscible polymer blends (Figures \ref{fig:predictions}c and \ref{fig:predictions}d) are illustrated in Figure \ref{fig:predictions}. The smoothness of the predictions across the composition range indicates that the machine learning model learned a smooth mapping between the fingerprint space and polymer properties. The experimental data points are in close agreement with the predictions, except for a few points in Figure \ref{fig:predictions}b that fall outside the shaded bands, which indicate the uncertainty of the predictions.

\section{Conclusion}
% P1
% speed, dataset, performance/validation, novel fingerprinting scheme (this a extension to previous scheme), advantage of blends (rapid and cost efficient development of new polymers, less chemical barriers for development), polymer chemical space (we add the polymer blend chemical space), miscibility
% beyond homopolymers, polymer informatics, beyond homopolymer informatics
% In this work, we developed a generalized framework for the prediction of homopolymer, copolymer, and polymer blend properties that completes polymer informatics. 
% P2
% future work: discover/screen (pre-screening before experimentation) new polymer blends, faster deployment of polymer blends, extend to more properties (this very easy), extending the fingerprint scheme, deployed at polymergenome.org, find compatibilizer to make immiscible polymer blends miscible, 
% first sentence should be short: more adjective (first-ever, exceptional performance,), emphasize universal 
% single
In this work, we developed a machine learning framework that is capable of predicting properties of homopolymers, copolymers, and polymer blends simultaneously with exceptional performance and surpasses previous works in terms of prediction accuracy and chemical domain for this property class \cite{Kuenneth2021}. Powered by multi-task predictors and a large dataset of 23,018 thermal data points, this framework enables the prediction of polymer properties that fall in a broad technologically relevant class. For the success of the model building, we designed a polymer blend fingerprinting scheme that extends and is based on homopolymer and copolymer fingerprints that we have used in the past. 

% This model can be used to select candidate materials for complex applications through high-throughput screening. Additionally, it can be used to predict miscible polymer blends. Our model is trained on polymer blends containing two constituents but it can easily be generalized to polymer blends containing more than two constituents using our fingerprinting scheme. 

There are several ways in which this work can be utilized to further expand the capabilities of polymer informatics in terms of accuracy and scope. Incorporating the impact of processing parameters and morphology in the case of polymer blends can lead to improvements in prediction accuracy and versatility. The conceptual ideas used in this work to set up the machine learning pipeline for thermal properties prediction for different polymer flavors can be extended to other properties (e.g., electronic and mechanical) when suitable data are available. Our approach can be adapted to different polymer classes, such as polymers with additives or polymer composites, by encoding features corresponding to the additive and fillers in the fingerprinting step. 

% A similar approach can be used to develop property-prediction models for other properties of polymer blends when suitable data is available. 
% \section{Code and Model Availability}
% The code is available at \url{https://github.com/Ramprasad-Group/copolymer_informatics}. The meta learner is openly available for use at \texttt{https://polymergenome.org}.

% \section{Author Contributions}
% C.K. designed, trained and evaluated the ML models. W.S. and C.K. collectively collected and curated the datapoints used in this study. The work was conceived and guided by R.R. All authors discussed results and commented on the manuscript. All authors have given approval to the final version of the manuscript.

\section{Methods}

\paragraph{Polymer Genome Fingerprints}
Polymer SMILES strings (e.g., \texttt{[*]CC[*]} for polyethylene) are converted to numerical fingerprint vectors using a hand-crafted fingerprinting scheme. For homopolymers/monomers, previous works have shown that effective fingerprinting involves numerically representing chemical and structural descriptors of the polymers at different length scales (atomic, block, and chain levels). \cite{DoanTran2020, Mannodi-Kanakkithodi2016} The benefits of these descriptors are that they have sufficient chemico-structural information to describe a wide range of physical and chemical attributes that control various polymer properties, they can distinguish two different monomers, and they are invariant to different specifications of the polymer SMILES strings of the same polymer.

\paragraph{Multi-task Models and Meta learner}
The polymer fingerprints along with the selector vector and thermal property values for homopolymers, copolymers, and polymer blends are used to train our predictive machine learning models. Before training, the thermal property values are curated using min-max normalization to scale them to [0,1]. For the classification models, polymer fingerprints, selector vectors, and the number of thermal property values for $T_\text{g}$ and $T_\text{m}$ of polymer blends are used for training. The machine learning architecture is shown in Figure \ref{fig:ml_arch}a, where the multi-task model is a concatenation-based conditioned multi-task deep neural network. This multi-task model is trained on \SI{80}{\%} of the dataset through five-fold cross validation (CV) and the remaining \SI{20}{\%} is utilized to train the meta learner. We utilized Tensorflow\cite{tensorflow} to implement all our models. Adam optimization along with stochastic weight averaging is used for updating the weights of the network. The learning rate is initially set at $10^{-3}$ and is changed in the later phases of training through the learning rate scheduler along with early stopping to prevent overfitting. Hyperparameters of our ML model that include the number of layers, number of neurons in each layer, initial learning rate, dropout rates, and the layer where the selector vector is concatenated were tuned using the Hyperband algorithm implemented in the Keras-Tuner \cite{kerastuner}. All values of the hyperparameters are present in Table S2 (in Supplementary Information). The machine learning architecture and the hyperparameter tuning steps are the same for the prediction and classification models.

The predicted thermal property values from the five-fold CV models are used as inputs to the meta learner as shown in Figure \ref{fig:ml_arch}b. The predictive meta learner is an ensemble model that predicts the final thermal property value from the five thermal property values of the five CV models. The meta learner for classifying miscibility is an ensemble model that predicts the final probabilities for miscibility of polymer blends. Both meta learners are trained on the remaining \SI{20}{\%} of the dataset that the cross validation models have never seen. Similar to the multi-task models, the hyperparameter optimization of the meta learner is achieved through the Hyperband algorithm implemented in the Keras-Tuner \cite{kerastuner}.

\section{Declarations}
\paragraph{Data Availability}
The datasets used in this study are from 
PolyInfo database \url{https://polymer.nims.go.jp/en/} (The copyrights of this database are owned by the National Institute for
Materials Science (NIMS)).

\paragraph{Code Avaibility}
The code is available at \url{https://github.com/Ramprasad-Group/Polymer_informatics_beyond_homopolymers}.

\paragraph{Author Contributions}
S.S.S curated the dataset and trained the ML models. C.K. provided assistance for training and evaluating the ML models. R.R. conceptualized the work and provided his guidance. All authors discussed the results and commented on the manuscript. 

\paragraph{Competing Interests}
The authors declare no competing financial interest.

\begin{acknowledgement}
This work is financially supported by the Office of Naval Research through a Multidisciplinary University Research Initiative (MURI) grant (N00014-17-1-2656). C.K. thanks the Alexander von Humboldt Foundation for financial support. 
\end{acknowledgement}

\bibliography{references_correct,references_old}

\end{document}